# COCI, the OpenCitations Index of Crossref open DOI-to-DOI citations

Ivan Heibi, ivan.heibi2@unibo.it, https://orcid.org/0000-0001-5366-5194
Digital Humanities Advanced Research Centre (DHARC), Department of Classical Philology and Italian Studies, University of Bologna, Bologna, Italy

Silvio Peroni, silvio.peroni@unibo.it, https://orcid.org/0000-0003-0530-4305
Digital Humanities Advanced Research Centre (DHARC), Department of Classical Philology and Italian Studies, University of Bologna, Bologna, Italy

David Shotton, david.shotton@oerc.ox.ac.uk, https://orcid.org/0000-0001-5506-523X
Oxford e-Research Centre, University of Oxford, Oxford, United Kingdom

**Corresponding author:** Silvio Peroni, silvio.peroni@unibo.it, +39 051 20 9 8576, via Zamboni 32, 40126 Bologna (BO), Italy

**Abstract**

In this paper, we present COCI, the OpenCitations Index of Crossref open DOI-to-DOI citations (http://opencitations.net/index/coci). COCI is the first open citation index created by OpenCitations, in which we have applied the concept of citations as first-class data entities, and it contains more than 445 million DOI-to-DOI citation links derived from the data available in Crossref. These citations are described using the Resource Description Framework (RDF) by means of the newly extended version of the OpenCitations Data Model (OCDM). We introduce the workflow we have developed for creating these data, and also show the additional services that facilitate the access to and querying of these data via different access points: a SPARQL endpoint, a REST API, bulk downloads, Web interfaces, and direct access to the citations via HTTP content negotiation. Finally, we present statistics regarding the use of COCI citation data, and we introduce several projects that have already started to use COCI data for different purposes.

**Article Highlights**

- COCI contains more than 445 million DOI-to-DOI citation links made available under a CC0 public domain waiver
- COCI uses an alternative richer view that regards citations as first-class data entities with accompanying properties
- Citation data in COCI can be accessed in a variety of ways including SPARQL endpoint, REST API, interfaces, and dumps

**Acknowledgements**

We gratefully acknowledge the financial support provided to us by the Alfred P. Sloan Foundation for the OpenCitations Enhancement Project (grant number G-2017-9800).

# Introduction

The availability of *open* scholarly citations (Peroni & Shotton 2018a) is a public good, of significant value to the academic community and the general public. In fact, citations not only serve as an acknowledgment medium (Newton, 1675), but also can be characterised topologically (by defining the connection graph between citing and cited entities and its evolution over time (Chawla 2017)), sociologically (such as for identifying unusual conduct within or elitist access paths to scientific research (Sugimoto et al. 2017)), quantitatively by creating citation-based metrics for evaluating the impact of an idea or a person (Schiermeier 2017), and 'financially' by defining the scholarly 'value' for a researcher within his/her own academic community (Molteni 2017). The Initiative for Open Citations (I4OC, https://i4oc.org) has dedicated the past two years to persuading publishers to provide open citation data by means of the Crossref platform (https://crossref.org), obtaining the release of the reference lists of more than 43 million articles (as of February 2019), and it is this change of behaviour by the majority of academic publishers that has permitted COCI to be created.

OpenCitations (http://opencitations.net) (Peroni & Shotton 2019b) is a scholarly infrastructure organization dedicated to open scholarship and the publication of open bibliographic and citation data by the use of Semantic Web (Linked Data) technologies, and is a founding member of I4OC. It has created and maintains the SPAR (Semantic Publishing and Referencing) Ontologies (http://www.sparontologies.net) (Peroni & Shotton 2018c) for encoding scholarly bibliographic and citation data in the Resource Description Framework (RDF) (Cyganiak, Wood & Krotzsch 2014), and has previously developed the OpenCitations Corpus (OCC) (Peroni, Shotton & Vitali 2017) of open downloadable bibliographic and citation data recorded in RDF.

In this paper, we introduce a new dataset made available a few months ago by OpenCitations, namely COCI, the OpenCitations Index of Crossref open DOI-to-DOI citations (https://w3id.org/oc/index/coci). This dataset, launched in July 2018, is the first of the indexes proposed by OpenCitations (https://w3id.org/oc/index), in which citations are exposed as first-class data entities with accompanying properties (i.e. individuals of the class `cito:Citation` as defined in CiTO (Peroni & Shotton 2012)) instead of being defined simply as relations among two bibliographic resources (via the property `cito:cites`). Currently COCI contains more than 445 million DOI-to-DOI citation links made available under a Creative Commons CC0 public domain waiver, that can be accessed and queried through a SPARQL endpoint (Harris & Seaborne, 2013), an HTTP REST API, by means of searching/browsing Web interfaces, by bulk download in different formats (CSV and N-Triples), or by direct access via HTTP content negotiation.

The rest of the paper is organized as follows. In 'Related works' we introduce some of the main RDF datasets containing scholarly bibliographic metadata and citations. In 'Indexing citations as first-class data entities', we provide some details on the rationale and the technologies used to describe citations as first-class data entities, which are the main foundations for the development of COCI. In 'COCI: ingestion workflow, data, and services', we present COCI, including the workflow process developed for ingesting and exposing the open citation data available, and other tools used for accessing these data. In 'Quantifying the use of COCI citation data', we show the scale of community uptake of COCI since its launch by means of quantitative statistics on the use of its related services and by listing existing projects that are using it for specific purposes. Finally, in 'Conclusions', we conclude the paper sketching out related and upcoming projects.

# Related works

We have noticed a recent growing interest within the Semantic Web community for creating and making available RDF ('Linked Data') datasets concerning the metadata of scholarly resources, particularly bibliographic resources. In this section, we briefly introduce some of the most relevant ones.

ScholarlyData (http://www.scholarlydata.org) (Nuzzolese et al. 2016) is a project that refactors the Semantic Web Dog Food so as to keep the dataset growing in good health. It uses the Conference Ontology, an improved version of the Semantic Web Conference Ontology, to describe metadata of documents (5,415, as of March 31, 2019), people (more than 1,100), and data about academic events (592) where such documents have been presented.

Another important source of bibliographic data in RDF is OpenAIRE (https://www.openaire.eu) (Alexiou et al. 2016). Created by funding from the European Union, its RDF dataset makes available data for around 34 million research products created in the context of around 2.5 million research projects.

While important, these aforementioned datasets do not provide citation links between publications as part of their RDF data. In contrast, the following datasets do include citation data as part of the information they make available.

In 2017, Springer Nature announced SciGraph (https://scigraph.springernature.com) (Hammond, Pasin & Theodoridis 2017), a Linked Open Data platform aggregating data sources from Springer Nature and other key partners managing scholarly domain data. It contains data about journal articles (around 8 millions, as of March 31, 2019) and book chapters (around 4.5 millions), including their related citations, and information on around 7 million people involved in the publishing process.

The OpenCitations Corpus (OCC, https://w3id.org/oc/corpus) (Peroni, Shotton & Vitali 2017) is a collection of open bibliographic and citation data created by ourselves, harvested from the open access literature available in PubMed Central. As of March 31, 2019, it contains information about almost 14 million citation links to more than 7.5 million cited bibliographic resources.

WikiCite (https://meta.wikimedia.org/wiki/WikiCite) is a proposal, with a related series of workshops, which aims at building a bibliographic database in Wikidata (Erxleben et al. 2014) to serve all Wikimedia projects. Currently Wikidata hosts (as of March 29, 2019) more than 170 million citations.

Biotea (https://biotea.github.io) (Garcia et al. 2018) is an RDF datasets containing information about some of the articles available in the Open Access subset of PubMed Central, that have been enhanced with specialized annotation pipelines. The last released dataset includes information extracted from 2,811 articles, including data on their citations.

Finally, Semantic Lancet (Bagnacani et al. 2014) proposes to build a dataset of scholarly publication metadata and citations (including the specification of the citation functions) starting from articles published by Elsevier. To date it includes bibliographic metadata, abstract and citations of 291 articles published in the Journal of Web Semantics.

## Indexing citations as first-class data entities

Citations are normally defined simply as links between published entities (from a citing entity to a cited entity). However, an alternative richer view is to regard each citation as a data entity in its own right, as illustrated in Fig. 1. This alternative approach permits us to endow a citation with descriptive properties, such as those introduced in Table 1[1].

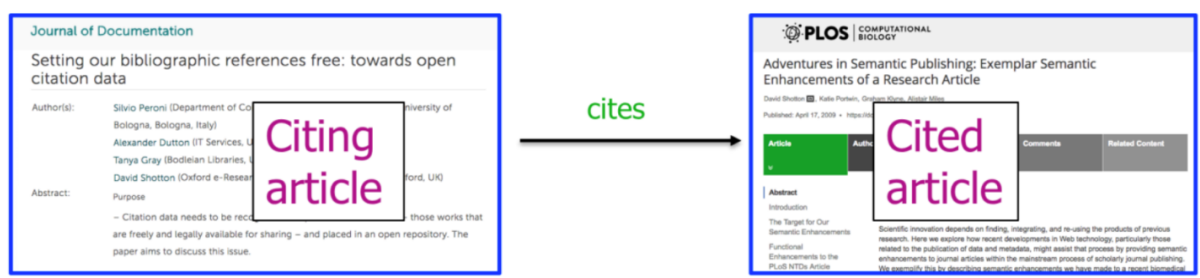

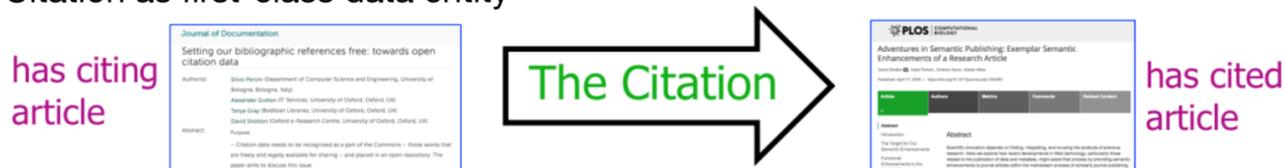

*Fig. 1* Two different ways of describing citations: as a relation between two bibliographic entities (top), or as an individual first-class data entity in its own right, where the citing entity and the cited entity are among its attributed data

The advantages of treating citations as first-class data entities are:

- all the information regarding each citation is available in one place, since such information is defined as attributes of the citation itself;
- citations become easier to describe, distinguish, count and process, and it becomes possible to distinguish separate in-text citation occurences within the citing entity to the cited entity, enabling one to count how many times, from which sections of the citing entity, and (in principle) for what purposes a particular cited entity is cited within the source paper;
- if available in aggregate, citations described in this manner are easier to analyse using bibliometric methods, for example to determine how citation time spans vary by discipline.

We have appropriately extended the OpenCitations Data Model (OCDM, http://opencitations.net/model) (Peroni & Shotton 2018a) so as to define each citation as a first-class entity in machine-readable manner. In particular, we have used the class `cito:Citation` defined in the revised and expanded Citation Typing Ontology (CiTO, http://purl.org/spar/cito) (Peroni & Shotton 2012), which is part of the SPAR Ontologies (Peroni & Shotton 2018b). This class allows us to define a permanent conceptual directional link from the citing bibliographic entity to a cited bibliographic entity, that can be accompanied by specific attributes, as introduced in Table 1.

---

[1] An in-depth description about the definition and use of citations as first-class data entities can be found at
https://opencitations.wordpress.com/2018/02/19/citations-as-first-class-data-entities-introduction/.

| Characteristic | Description |
| --- | --- |
| citing entity | The bibliographic entity which acts as source for the citation. |
| cited entity | The bibliographic entity which acts as target for the citation. |
| citation creation date | The date on which the citation was created. This has the same numerical value as the publication date of the citing bibliographic resource, but is a property of the citation itself. When combined with the citation time span, it permits that citation to be located in history. |
| citation timespan | The temporal characteristic of a citation, namely the interval between the publication date of the cited entity and the publication date of the citing entity. |
| type | A classification of the citation according to particular dimensions, e.g. whether or not it is a self-citation. |

*Table 1: List of characteristics that can be associated with a citation when it is described as first-class data entity.*

So as to identify each citation precisely, when described as a first-class data entity and included in an open dataset, we have also developed the Open Citation Identifier (OCI) (Peroni & Shotton 2019a), which is a new globally unique persistent identifier for citations. OCIs are registered in the Identifiers.org platform (https://identifiers.org/oci) and recognized as persistent identifiers for citations by the EU FREYA Project (https://www.project-freya.eu) (Ferguson et al. 2018). Each OCI has a simple structure: the lower-case letters "oci" followed by a colon, followed by two sequences of numerals separated by a dash, where the first sequence is the identifier for the citing bibliographic resource and the second sequence is the identifier for the cited bibliographic resource. For example, `oci:0301-03018` is a valid OCI for a citation defined within the OpenCitations Corpus, while `oci:02001010806360107050663080702026306630509-02001010806360107050663080702026305630301` is a valid OCI for a citation included in Crossref. It is worth mentioning that OCIs are not opaque identifiers, since they explicitly encode directional relationships between identified citing and cited entities, the provenance of the citation, i.e. the database that contains it, and the type of identifiers used in that database to identify the citing and cited entities. In addition, we have created the Open Citation Identifier Resolution Service (http://opencitations.net/oci), which is a resolution service for OCIs based on the Python application `oci.py` available at https://github.com/opencitations/oci. Given a valid OCI as input, this resolution service is able to retrieve citation data and present it in RDF or in Scholix, JSON or CSV formats. A more detailed explanation of OCIs and related material is available in (Peroni & Shotton 2019a).

At OpenCitations, we define an open citation index as a dataset containing citations that complies with the following requirements:

- the citations contained are all open, according to the definition provided in (Peroni & Shotton 2018a);
- the citations are treated as first-class data entities;
- each citation is identified by an Open Citation Identifier (OCI) (Peroni & Shotton 2019a);
- the citation data are recorded in RDF according to the OpenCitations Data Model (OCDM) (Peroni & Shotton 2018b), where the OCI of a citation is embedded in the IRI defining it in RDF;
- the attributes for citations shown in Table 1 are defined.

## COCI: ingestion workflow, data, and services

COCI, the OpenCitations Index of Crossref open DOI-to-DOI references, is the first citation index to be published by OpenCitations, in which we have applied the concept of citations as first-class data entities, introduced in the previous section, to index the contents of one of the major open databases of scholarly citation information, namely Crossref (https://crossref.org), and to render and make available this information in machine-readable RDF under a CC0 waiver. Crossref contains metadata about publications (mainly academic journal articles) that are identified using Digital Object Identifiers (DOIs). Out of more than 100 million publications recorded in Crossref, Crossref now stores the reference lists of more than 43 million of them deposited by the publishers. Many of these references are to other publications bearing DOIs that are also described in Crossref, while others are to publications that lack DOIs and do not have Crossref descriptions. Crossref organises such publications with associated reference lists according to three categories: *closed*, *limited* and *open*. These categories refer, respectively, to publications for which the reference lists are not visible to anyone outside the Crossref Cited-by membership, are visible only to them and to Crossref Metadata Plus members, or are visible to and open for re-use by all users[2].

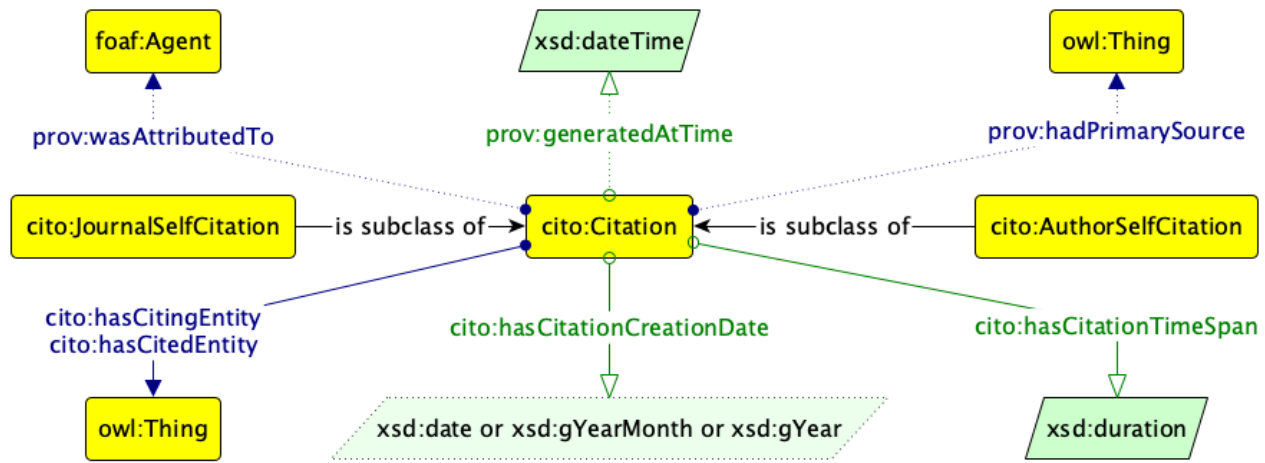

***Fig. 2*** *The diagram of the data model adopted to define the new class cito:Citaton for describing citations as first-class data entities, which forms part of the OpenCitations Data Model. This model uses terms from the Citation Typing Ontology (CiTO, http://purl.org/spar/cito) for describing the data, and from the Provenance Ontology (PROV-O, http://www.w3.org/ns/prov) to define the citation's provenance*

Followed the first release of COCI on June 4, 2018, the most recent version of COCI, released on November 12, 2018, contains more that 445 million DOI-to-DOI citations that are included in the *open* and the *limited* datasets of Crossref reference data[3]. All the citation data in COCI and their provenance information, described according the Graffoo diagram (Falco et al. 2014) presented in Fig. 2, are released under a CC0 waiver, and are compliant with the FAIR data principles (Wilkinson et al. 2016).

In the following subsections we introduce the ingestion workflow developed for creating COCI, we provides some figures on the citations it contains, and we list the resources and services we have made available to permit access to and querying of the dataset.

**Ingestion workflow**

We processed all the data included in the October 2018 JSON dump of Crossref data, available to all the Crossref Metadata Plus members. The ingestion workflow, summarised in Fig. 3, was organised in four distinct phases, and all the related scripts developed and used for this purpose are released as open source code according to the ISC License and downloadable from the official GitHub repository of COCI at https://github.com/opencitations/coci.

---

[2]Additional information on this classification of Crossref reference lists is available at https://www.crossref.org/reference-distribution/.
[3]We have access to the *limited* dataset since we are members of the Crossref Metadata Plus plan.

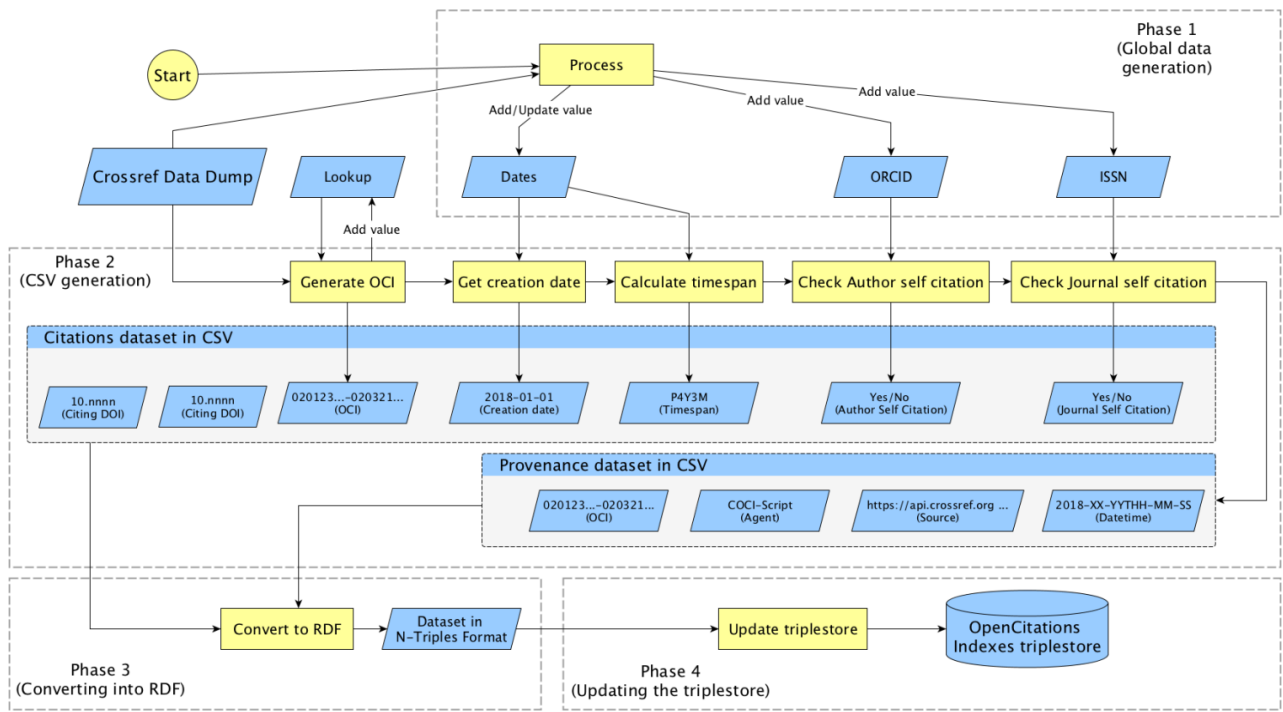

*Fig. 3 A flowchart scheme describing the workflow to build COCI. It is divided in four phases: (1) global data generation, (2) CSV generation, (3) conversion into RDF, and (4) updating the triplestore*

**Phase 1: global data generation.** We parse and process the entire Crossref bibliographic database to extract all the publications having a DOI and their available list of references. Through this process three datasets are generated, which are used in the next phase:

- *Dates*, the publication dates of all the bibliographic entities in Crossref and of all their references if they explicitly specify a DOI and a publication date as structured data – e.g. see the fields "DOI" and "year" in the array "reference" in https://api.crossref.org/works/10.1007/978-3-030-00668-6_8. Where the same DOI is encountered multiple times, e.g. as a proper item indexed in Crossref and also as a reference in the reference list of another article deposited in the Crossref, we use the full publication date defined in the indexed item.
- *ISSN*: the ISSN (if any) and publication type ("journal-article", "book-chapter", etc.) of each bibliographic entity identified by a DOI indexed in Crossref.
- *ORCID*: the ORCIDs (if any) associated with the authors of each bibliographic entity identified by a DOI indexed in Crossref.

**Phase 2: CSV generation.** We generate a CSV file within which each row represents a particular citation between a citing entity and a cited entity according to the data available in the Crossref dump, by looking at the DOI identifying the citing entity and all the DOIs specified in the reference list of such a citing entity according to the Crossref data. In particular, we execute the following four steps for each citation identified:

1. We generate the OCI for the citation by encoding the DOIs of the citing and cited entities into numerical sequences using the lookup table available at https://github.com/opencitations/oci/blob/master/lookup.csv, which are prefixed by the supplier prefix "020" to indicate Crossref as the source of the citation.
2. We retrieve the publication date of the citing entity from the *Dates* dataset and assign it as citation creation date.
3. We retrieve the publication date of the cited entity (from the *Dates* dataset) and we use it, together with the publication date of the citing entity retrieved in the previous step, to calculate the citation timespan.
4. We use the data contained in the *ISSN* and *ORCID* datasets to establish whether the citing and cited entity have been published in the same journal and/or have at least one author in common, and in these cases we assign the appropriate self-citation type(s) to the citation.

Simultaneously with the creation of the CSV file of citation data, we generate a second CSV file containing the provenance information for each citation (identified by its OCI generated in the aforementioned Step 1). These provenance data include the agent responsible for the generation of the citation, the Crossref API call that refers to the data of the citing bibliographic entity containing the reference used to create the citation, and the creation date of the citation.

**Phase 3: converting into RDF.** The CSV files generated in the previous phase are then converted into RDF according to the N-Triples format, following the OWL model introduced in Fig. 2, where the DOIs of the citing and cited entities become DOI URLs starting with "http://dx.doi.org/"[4], while the IRI of the citation includes its OCI (without the "oci:" prefix), as illustrated in the example given in the previous section.

**Phase 4: updating the triplestore.** The final RDF files generated in Phase 3 are used to update the triplestore used for the OpenCitations Indexes.

## Data

COCI was first created and released on July 4, 2018, and most recently updated on November 12, 2018. Currently, it contains 445,826,118 citations between 46,534,705 bibliographic entities. These are stored by means of 2,259,134,894 RDF statements (around 5 RDF statements per citation) for describing the citation data, and 1,337,478,354 RDF statements (3 statements per citation) for describing the related provenance information. Of the citations stored, 29,755,045 (6.7%) are journal self-citations, while 250,991 (0.06%) are author self-citations. The number of identified author self-citations, based on author ORCIDs, is a significant underestimate of the true number, mainly due to the sparsity of the data concerning the ORCID author identifiers within the Crossref database. Journal entities (i.e. journals, volumes, issues, and articles) are the most common type of bibliographic entity cited, with over 420 million citations.

We also classify the cited documents according to their publishers – Table 2 shows the ten top publishers of citing and cited documents, calculated by looking at the DOI prefixes of the entities involved in each citation. As we can see, Elsevier is by far the publisher having the majority of cited documents. It is also the largest publisher that is **not** participating in the Initiative for Open Citations by **not** making its publications' reference lists open at Crossref – which is highlighted by the very limited amount of outgoing citations recorded in COCI. Its present refusal to open its article reference lists in Crossref, contrary to the practice of most of the major scholarly publishers, is contributing significantly to the invisibility of Elsevier's own publications within the corpora of open citation data such as COCI that are increasingly being used by the scholarly community for discovery, citation network visualization and bibliometric analysis, as we introduce below in the section entitled Quantifying the use of COCI citation data.

| **Publisher** | **Outgoing citations** | **Incoming citations** |
|---|---|---|
| Springer Nature | 79,860,827 | 52,257,862 |
| Wiley | 76,819,685 | 48,174,542 |
| Informa UK Limited | 41,433,917 | 14,975,989 |
| *Institute of Electrical and Electronics Engineers (IEEE)* | 30,114,985 | 20,940,703 |
| SAGE Publications | 15,933,805 | 7,915,082 |
| American Physical Society (APS) | 15,729,297 | 16,065,862 |
| AIP Publishing | 10,130,022 | 8,455,097 |
| *Ovid Technologies (Wolters Kluwer Health)* | 9,971,274 | 12,840,293 |
| Oxford University Press (OUP) | 9,891,000 | 11,466,659 |
| *Elsevier* | 2,853,739 | 96,310,027 |

*Table 2. A classification of the COCI citations according to the publishers of the cited ("incoming citations") and citing ("outgoing citations") documents. The table shows the top ten publishers by the overall amount of incoming and outgoing to/from their published works. Those publishers shown in italics are not participating in the Initiative for Open Citations by making their publications' reference lists open at Crossref – see https://i4oc.org for additional information.*

## Resources and services

The citation data in COCI can be accessed in a variety of convenient ways, listed as follows.

**Open Citation Index SPARQL endpoint.** We have made available a SPARQL endpoint for all the indexes released by OpenCitations, including COCI, which is available at https://w3id.org/oc/index/sparql. When accessed with a browser, it shows a SPARQL endpoint editor GUI generated with YASGUI (Rietveld & Hoekstra 2017). Of course, this SPARQL endpoint can additionally be queried using the REST HTTP protocol, e.g. via *curl*. In order to access COCI data, the graph https://w3id.org/oc/index/coci/ must be specified in the SPARQL query.

---
[4]We are aware that the current practice for DOI URLs is to use the base "https://doi.org/" instead of "http://dx.doi.org/". However, when one tries to resolve a DOI URL owned by Crossref by specifying an RDF format in the accept header of the request, the bibliographic entity is actually defined using the old URL structure starting with "http://dx.doi.org/". For this reason, since COCI is derived entirely from Crossref data, we decided to stay with the approach currently used by Crossref.

**COCI REST API.** Citation data in COCI can be retrieved by using the COCI REST API, available at https://w3id.org/oc/index/coci/api/v1. The rationale of making available a REST API – implemented by means of RAMOSE, the Restful API Manager Over SPARQL Endpoints (https://github.com/opencitations/ramose) – in addition to the SPARQL endpoint was to provide convenient access to the citation data included in COCI for Web developers and users who are not necessarily experts in Semantic Web technologies. The COCI REST API makes available four operations, that will retrieve either (a) the citation data for all the outgoing references of a given DOI (operation: *references*), or (b) the citation data for all the incoming citations received by a given DOI (operation: *citations*), or (c) the citation data for the citation identified by an OCI (operation: *citation*), or (d) the metadata for the article(s) identified by the specified DOI or DOIs (operation: *metadata*). It is worth mentioning that the latter operation strictly depends on live API calls to external services to gather the metadata of the requested articles, such as the title, the authors, and the journal name, that are not explicitly included within the OpenCitations Index triplestore.

**Searching and browsing interfaces.** We have additionally developed a user-friendly text search interface (https://w3id.org/oc/index/search), and a browsing interface (e.g. https://w3id.org/oc/index/browser/coci/ci/02001010806360107050663080702026306630509-02001010806360107050663080702026305630301), that can be used to search citation data in all the OpenCitations Indexes, including COCI, and to visualise and browse them, respectively. These two interfaces have been developed by means of OSCAR, the OpenCitations RDF Search Application (https://github.com/opencitations/oscar) (Heibi, Peroni & Shotton 2019b), and LUCINDA, the OpenCitations RDF Resource Browser (https://github.com/opencitations/lucinda), that provide a configurable layer over SPARQL endpoints that permit one easily to create Web interfaces for querying and visualising the results of SPARQL queries.

**Data dumps.** All the citation data and provenance information in COCI are available as dumps stored in Figshare (https://figshare.com) in both CSV and N-Triples formats, while a dump of the whole triplestore is available on The Internet Archive (https://archive.org). The links to these dumps are available on the download page of the OpenCitations website (http://opencitations.net/download#coci).

**Direct HTTP access.** All the citation data in COCI can be accessed directly by means of the HTTP IRIs of the stored resources (via content negotiation, e.g. https://w3id.org/oc/index/coci/ci/02001010806360107050663080702026306630509-02001010806360107050663080702026305630301).

## Quantifying the use of COCI citation data

In the past months, we have monitored the accesses to COCI data since its launch in July 2018. The statistics and graphics we show in this section highlight two different aspects: the quantification of the use of COCI data – and related services – and the community uptake, i.e. the use of COCI data for specific reuses within cross-community projects and studies. All the data of the charts described in this section are freely available for download from Figshare (Heibi, Peroni & Shotton 2019c).

### Quantitative analysis

Fig. 4 shows the number of accesses made between July 2018 and February 2019 (inclusive) to the various COCI services described above – the search/browse interfaces, the REST API, SPARQL queries, and others (e.g. direct HTTP access to particular citations and visits to COCI webpages in the OpenCitations website). We have excluded from all these counts all accesses made by automated agents and bots. As shown, the REST API is, by far, the most used service, with extensive usage recorded in the last four months shown, following the announcement of the second release of COCI. This is reasonable, considering that the REST API has been developed exactly for accommodating the needs of generic Web users and developers, including (and in particular) those who are not expert in Semantic Web technologies. There is just one exception in November 2018, where the SPARQL endpoint was used to retrieve quite a large amount of citation data. After further investigation, we noticed a large proportion of the SPARQL calls were coming from a single source (according to the IP data stored in our log), which probably collected citation data for a specific set of entities.

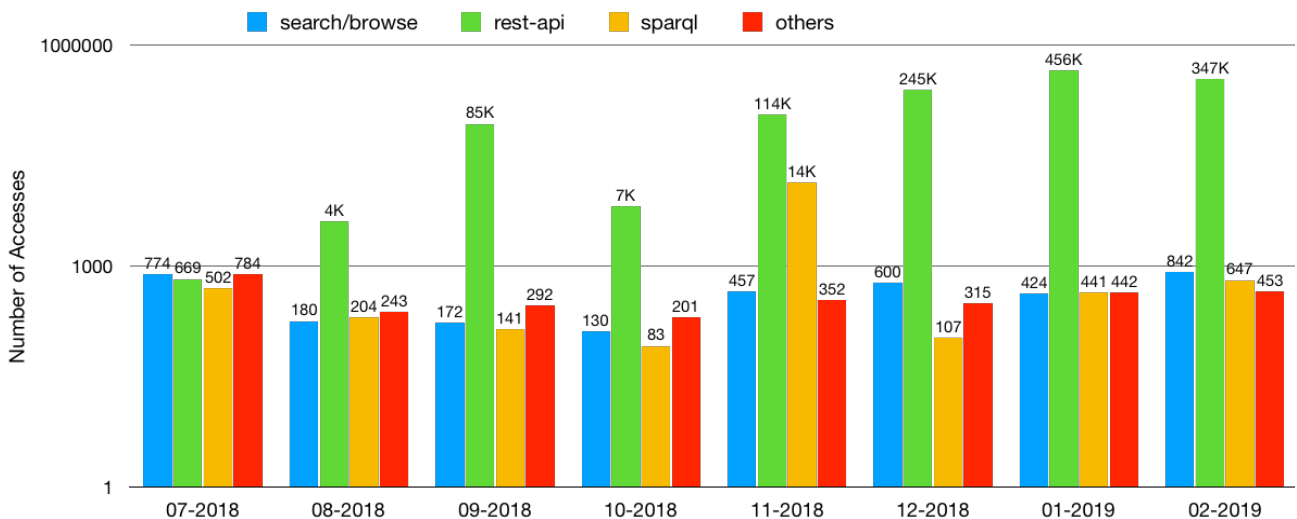
*Fig. 4* The number of accesses to COCI-related services since July 2018 to February 2019. The scale used in the y-axis is logarithmic.

Fig. 5 shows a particular cut of the figures given in Fig. 4, which focuses on the REST API accesses only. In particular, we analysed which operations of the API were used the most. According to these figures, the most used operation is *metadata* (which was first introduced in the API in August 2018) which allows one to retrieve all the metadata describing certain publications. In contrast to the other API operations that provide metadata relating to a single citation or citation information relating to a single publication, this metadata search accepts one or more DOIs as input, thus providing bibliographic metadata on one or more publications. The least used operation was *citation*, which allows one to retrieve citation data given an OCI, which should not be surprising, considering the currently limited knowledge of this new identifier system for citations.

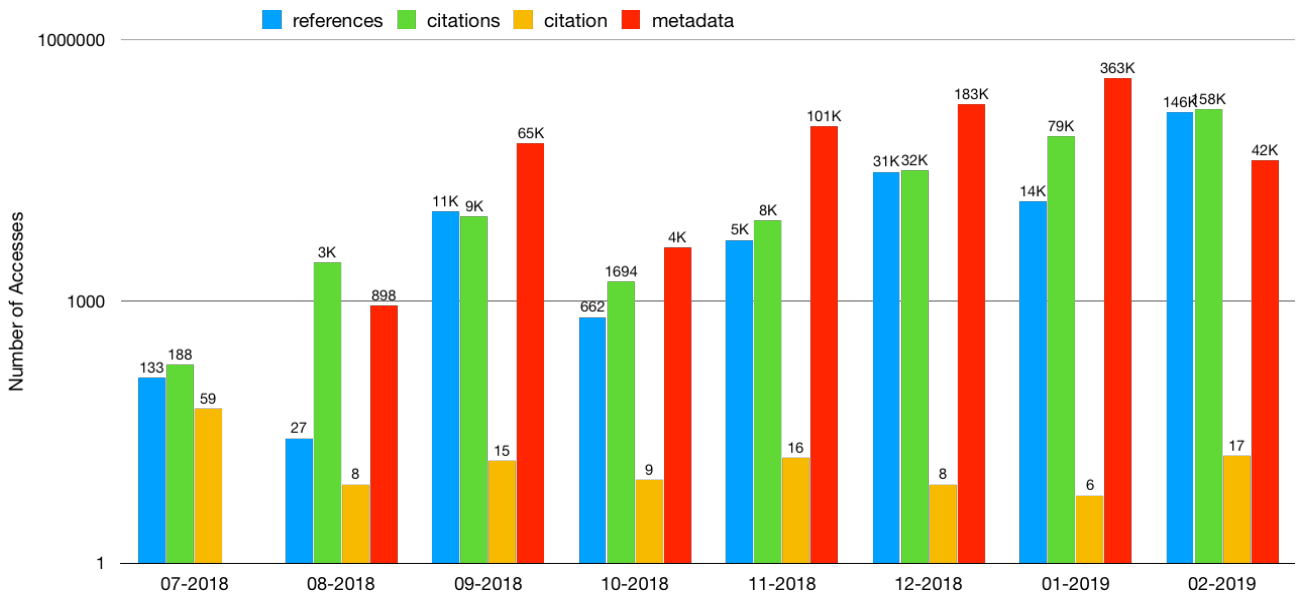
*Fig. 5* The number of access made to each different COCI REST-API operation since the release of COCI on July 2018. Classified into 4 categories (requested resource): 'references', 'citations', 'citation', and 'metadata', as defined in the text. Note again the logarithmic scale of the y-axis.

In addition, we have also retrieved data about the views and downloads (as of March 29, 2019) of all the dumps uploaded to Figshare and to the Internet Archive. The CSV data dump received 1,321 views and 454 downloads, followed by the N-Triples data dump with 316 views and 93 downloads. The CSV provenance information dump has 166 views and 127 downloads, while the N-Triples provenance information dump had 95 views and 34 downloads. Finally, the least accessed dump was that of the entire triplestore available in the Internet Archive, uploaded for the very first time in November 2018, that had only 3 views.

## Community uptake

The data in COCI has been already used in various projects and initiatives. In this section, we list all the tools and studies doing this of which we are aware.

VOSviewer (http://www.vosviewer.com) (van Eck & Waltman 2009) is a software tool, developed at the Leiden University's Centre for Science and Technology Studies (CWTS), for constructing and visualizing bibliometric networks, which may include journals, researchers, or individual publications, and may be constructed based on citation, bibliographic coupling, co-citation, and co-authorship relations. Starting from version 1.6.10 (released on January 10, 2019), VOSviewer can now directly use citation data stored in COCI, retrieved by means of the COCI REST API.

Citation Gecko (http://citationgecko.com) is a novel literature mapping tool that allows one to map a research citation network using some initial seed articles. Citation Gecko is able to leverage citation links between seed papers and other papers to highlight papers of possible interest to the user, for which it uses COCI data (accessed via the REST API) to generate the citation network.

OCI Graphe (https://dossier-ng.univ-st-etienne.fr/scd/www/oci/OCI_graphe_accueil.html) is a Web tool that allows one to search articles by means of the COCI REST API, that are then visualised in a graph showing citations to the retrieved articles. It enriches this visualisation by adding additional information about the publication venues, publication dates, and other related metadata.

Zotero (Ahmed & Al Dhubaib 2011) is a free, easy-to-use tool to help users collect, organize, cite, and share research. Recently, the Open Citations Plugin for Zotero (https://github.com/zuphilip/zotero-open-citations) has been released, which allows users to retrieve open citation data extracted from COCI (via its REST API) for one or more articles included in a Zotero library.

COCI data, downloaded from the CSV dump available on Figshare, have been also used in at least two bibliometric studies. In particular, during the LIS Bibliometrics 2019 Event, Stephen Pearson presented a study (https://blog.research-plus.library.manchester.ac.uk/2019/03/04/using-open-citation-data-to-identify-new-research-opportunities/) run on publications by scholars at the University of Manchester which used COCI to retrieve citations between these publications so as to investigate possible cross-discipline and cross-department potential collaborations. Similarly, COCI data were used to conduct an experiment on the latest Italian Scientific Habilitation (Di Iorio, Peroni & Poggi 2019) (the national exercise that evaluates whether a scholar is appropriate to receive an Associate/Full Professorship position in an Italian university), which aimed at trying to replicate part of the outcomes of this evaluation exercise for the Computer Science research field by using only open scholarly data, including the citations available in COCI, rather than citation data from subscription services. Finally, COCI has also been used to explore the roles of books in scholarly communication (Zhu et al. 2019).

## Conclusions

In this paper, we have introduced COCI, the OpenCitations Index of Crossref open DOI-to-DOI citations. After an initial introduction of the notion of citations as first-class data entities, we have presented the ingestion workflow that has been implemented to create COCI, have detailed the data COCI contains, and have described the various services and resources that we have made available to access COCI data. Finally, we have presented some statistics about the use of COCI data, and have mentioned the tools and studies that have adopted COCI in recent months.

COCI is just the first open citations index that OpenCitations will make available. Using the experience we have gathered by creating it, we now plan the release of additional indexes, so as to extend the coverage of open citations available through the OpenCitations infrastructure. The first of these, recently released, is CROCI (https://w3id.org/oc/index/croci) (Heibi, Peroni & Shotton 2019a), the Crowdsourced Open Citations Index, which contains citations deposited by individuals. CROCI is designed to permit scholars proactively to fill the "open citations gap" in COCI resulting from four causes: (a) the failure of many publishers using Crossref DOIs to deposit reference lists of their publications at Crossref, (b) the failure of some publishers that *do* deposit their reference lists to make these reference lists open, in accordance with the recommendations of the Initiative for Open Citations; (c) the absence, in the October 2018 Crossref data dump, from ~11% of Crossref reference metadata of the DOIs for cited articles which in fact have been assigned DOIs (https://www.crossref.org/blog/underreporting-of-matched-references-in-crossref-metadata/) a problem that Crossref has recently rectified, so that citations to those articles will be included in the next update of COCI; and (d) the existence of citations to published entities that lack Crossref DOIs. In the near future, we plan to extend the number of indexes by harvesting citations from other open datasets including Wikidata (https://www.wikidata.org), DataCite (https://datacite.org), and Dryad (https://datadryad.org). In addition, we plan to extend and generalise the current software developed for COCI, so as to facilitate more frequent updates of the indexes.